%% file: MAIN.tex
\documentclass{ifacconf}

\usepackage{graphicx}      
\usepackage{natbib}        

\input{workPackages}

\begin{document}
\begin{frontmatter}

\title{%
\mbox{\begin{tabular}{c}Self-Healing Hybrid Control as a Proxy for Detection and \\
Mitigation of Sensor Attacks in Cooperative Driving\thanksref{footnoteinfo}\end{tabular}}%
} 

\thanks[footnoteinfo]{The research leading to these results has received funding from the European Union’s Horizon Europe programme under grant agreement No 101069748 – SELFY project.}

\author[First]{M.R. Huisman} 
\author[First,Second]{C.G. Murguia} 
\author[First]{E. Lefeber}
\author[First]{N. v/d Wouw}

\address[First]{Department of Mechanical Engineering, Eindhoven University of Technology, The Netherlands (e-mail: [m.r.huisman; c.g.Murguia; a.a.j.lefeber; n.v.d.wouw]@tue.nl).}
\address[Second]{murguia\_rendon@sutd.edu.sg, Singapore University of Technology and Design, Engineering Systems Design Pillar.}
\begin{abstract}                
We propose a real-time hybrid controller scheme to detect and mitigate False-Data Injection (FDI) attacks on Cooperative Adaptive Cruise Control (CACC). Our method uses sensor redundancy to create equivalent controller realizations, each driven by distinct sensor subsets but producing identical control inputs when no attack occurs. By comparing control signals and measurements via majority voting, the scheme identifies compromised sensors in real-time and switches to a healthy controller, even under unconstrained attacker switching. The hybrid controller utilizes attack-dependent flow and jump sets, and resets the states of compromised controllers, resulting in a self-healing architecture. Simulation results demonstrate the effectiveness of this approach.\vspace{-3mm}
\end{abstract}


\end{frontmatter}


\input{1.Introduction}

\input{2.Preliminaries}

\input{3.ModelDescription}

\input{4.Approach}

\input{5.Results}

\input{6.ConclusionFuturework}

\section*{DECLARATION OF GENERATIVE AI AND AI-ASSISTED TECHNOLOGIES IN THE WRITING PROCESS} \vspace{-3mm}
During the preparation of this work, the author(s) used ChatGPT to check grammar. After using this ChatGPT, the author(s) reviewed and edited the content as needed and take(s) full responsibility for the content of the publication.

\bibliography{ifacconf}     
\end{document}

%% file: workPackages.tex
\usepackage{epsfig} 
\usepackage{times} 
\usepackage{amsmath} 
\usepackage{amssymb}  
\usepackage{mathtools}
\usepackage{arydshln}
\usepackage{comment}
\usepackage{xcolor}
\usepackage{array}
\usepackage{diagbox}
\usepackage{booktabs} 
\usepackage{makecell} 
\usepackage{url}

\usepackage{tikz}
\usepackage{mathdots}
\usepackage{yhmath}
\usepackage{cancel}
\usepackage{siunitx}
\usepackage{multirow}
\usepackage{textcomp, gensymb}
\usepackage{tabularx}
\usepackage{extarrows}
\usetikzlibrary{fadings}
\usetikzlibrary{patterns}
\usetikzlibrary{shadows.blur}
\usetikzlibrary{shapes}
\usepackage{adjustbox}

\newtheorem{definition}{Definition}
\newtheorem{assumption}{Assumption}

\newcommand{\vim}{v_{i-1}}
\newcommand{\aim}{a_{i-1}}

\newcommand{\uim}{u_{i-1}}
\newcommand{\imo}{_{i-1}}

%% file: 1.Introduction.tex
\section{INTRODUCTION} \vspace{-3mm}
Cooperative Adaptive Cruise Control (CACC) is a well-explored technology within Connected and Automated Vehicles (CAVs) with the potential to improve transportation efficiency; see \cite{ploeg_design_2011}. Since CACC requires vehicle-to-vehicle (V2V) communication, \cite{el-rewini_cybersecurity_2020} highlights that network access points are exposed that adversaries can exploit for cyberattacks. To counter these attacks, techniques have been developed for attack prevention and detection; see \cite{ju_survey_2022}, \cite{sun_survey_2022}, \cite{chowdhury_attacks_2020}. 

Hybrid switching strategies have emerged as a promising tool to enhance security. While state estimation techniques aim to maintain accurate state estimates under sensor attacks (\cite{chong_observability_2015}, \cite{ding_secure_2021}, \cite{niazi_resilient_2023}), hybrid approaches employ a bank of observers or controllers that operate in parallel and actively switch between them. For instance, \cite{petri_towards_2022} studied such strategies to enhance state estimation performance (without addressing attacks), whereas~\cite{lu_secure_2017} applied such strategies to detect and mitigate sensor attacks, although with dwell-time constraints on the attacker’s switching frequency. Controller-switching is used only as a fallback mechanism after detection, rather than actively detecting attacks~(\cite{van_der_heijden_analyzing_2017}).

Despite their effectiveness, observer-based methods rely on satisfying strict observability criteria, which, given limited measurements, remains a significant challenge (\cite{ju_survey_2022}). Constructing multiple observers can also result in computational complexity, as the number of observer states quickly increases~(\cite{mao_computational_2022}). Furthermore, switching control schemes often sacrifice desired closed-loop performance to ensure security.

In this paper, equivalent controller realizations, introduced in~\cite{huisman_optimal_2024}, are integrated into a hybrid control scheme to detect and mitigate False-Data Injection (FDI) attacks in a dynamic CACC scheme. By reformulating the CACC scheme, two distinct controller realizations are derived that require as few sensors as possible while preserving closed-loop behavior. Each controller realization uses a distinct subset of sensors, providing flexibility in achieving the necessary sensor redundancy to be secure against attacks. 

We define discrete operating modes, each corresponding to a distinct attack scenario. The properties of the two controller realizations allow the hybrid scheme to switch between them. When the attacker stops or switches to a different sensor, the controller state of the previously compromised controller is reset using a healthy one. This reset prevents lingering attack effects and ensures the previously compromised controller can function correctly alongside the other healthy controllers. By switching between the different modes, the scheme both detects and mitigates FDI attacks and selects a healthy control input, preventing the attack from disrupting the platooning behavior.

The hybrid controller enables real-time detection and mitigation without relying on dwell-time assumptions, allowing the attacker to switch attacks instantaneously. To support our claims, we conduct a simulation study of a vehicle platoon using the hybrid control scheme implemented with the Hybrid Equations Toolbox in MATLAB Simulink (\cite{sanfelice_toolbox_2013}). Our results demonstrate that a healthy control input can always be selected even if the attacker switches. 

The structure of the paper is as follows. Section~\ref{sec:Preliminaries} introduces preliminary results. Section~\ref{sec:ProblemSetting} outlines the problem setting, including the platooning dynamics and the CACC scheme. Section~\ref{sec:controller_realization} presents the two distinct controller realizations, and Section~\ref{sec:Hybrid_Control_Scheme} details the hybrid control scheme. Section~\ref{sec:Results} reports simulation results, and Section~\ref{sec:Conclusion} concludes the paper.

%% file: 2.Preliminaries.tex



\section{Notation and Definitions} \vspace{-3mm} \label{sec:Preliminaries}
The symbol $\mathbb{R}$ stands for the real numbers, $\mathbb{R}_{>0}$ ($\mathbb{R}_{\geq 0}$) denotes the set of positive (non-negative) real numbers. The symbol $\mathbb{N}$ denotes the set of natural numbers, including zero. The $n \! \times \! m$ matrix composed of only zeros is denoted by $\mathbf{0}_{n \times m}$, or $\mathbf{0}$ when its dimension is clear. Time dependencies of signals are often omitted for simplicity of notation.

\begin{definition}[Hybrid Dynamical System]\emph{\cite{goebel_hybrid_2012}}\label{DefHybridSystem}
Consider the hybrid dynamical system:
\begin{align} \label{eq:HybridGeneral}
    \left\{ \begin{aligned}
    \dot{x} &= f(x,u), \text{ if }(x,u)\in \mathcal C, \\
        x^+ &= g(x,u), \text{ if }(x,u)\in \mathcal D,
    \end{aligned} \right.
\end{align}
where $x\in \mathbb R^{n_x}$ is the state vector, $\dot x$ represents its continuous evolution, and $x^+$ denotes the state after an instantaneous jump. The input vector is given by $u \in \mathbb R^{n_u}$. The function $f(x,u)$ defines the flow map and $g(x,u)$ represents the jump map, with their corresponding flow set $\mathcal C$, and jump set $\mathcal D$, respectively. The solutions of \eqref{eq:HybridGeneral} evolve over a hybrid time domain $(t,j) \in \mathbb E$ where $t$ represents continuous time, and $j$ represents the discrete jump counter. The subset $\mathbb E \subset \mathbb R_{>0} \times \mathbb N$ is called a \textit{compact hybrid time domain} if $\mathbb E = \bigcup^J_{j=0}([t_j, t_{j+1}],j)$ for some finite sequence of times $0 = t_0 \leq ... \leq t_{J+1}$. We say $\mathbb E$ is a \textit{hybrid time domain} if, for each $(T,J)\in \mathbb E$, the intersection $\mathbb E \bigcap([0,T] \times \{0,...,J\})$ is a compact hybrid time domain.
\end{definition}

%% file: 3.ModelDescription.tex
\section{Problem Setting} \label{sec:ProblemSetting} \vspace{-3mm}
\input{3_ProblemSetting/3.1_ProblemSettingV2}

%% file: 3_ProblemSetting/3.1_ProblemSettingV2.tex
\begin{figure}[bt]\centering
		\includegraphics[width=\linewidth]{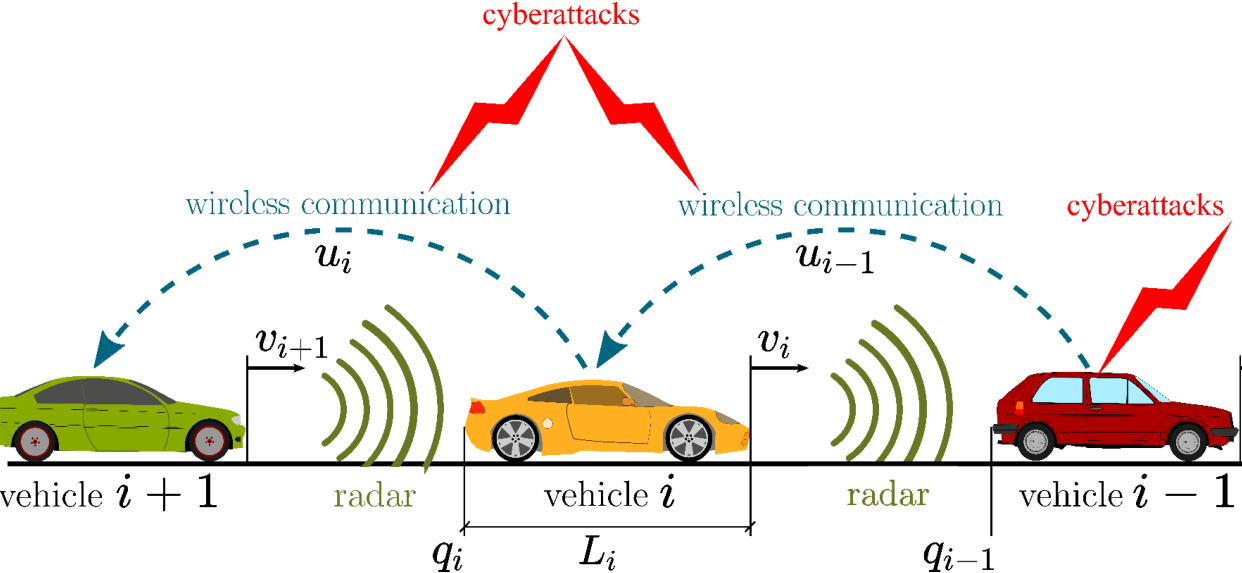} \vspace{-1mm}
		\caption{Vehicle platoon with CACC. Each vehicle has onboard sensors (e.g., LiDARs, cameras, and velocity/acceleration sensors) that may be subject to FDI attacks.}
		\centering
		\label{fig:Platoon}
  \vspace{0mm}
\end{figure}

Consider a homogeneous platoon of $m$ vehicles, as illustrated in Fig.~\ref{fig:Platoon}. The vehicles are indexed by $i=1,...,m,$ where $i=1$ indicates the lead vehicle. To describe the platooning behavior, we adopt a linear platooning model \cite{ploeg_design_2011}:
\begin{align} \label{eq:SD_PlatoonDynamics}
    \begin{bmatrix}
        \dot{v}\imo \\ \dot{a}\imo \\ \dot{d}_i \\ \dot{v}_i \\\dot{a}_i
    \end{bmatrix}
    = 
    \begin{bmatrix}
        a\imo \\ -\tfrac{1}{\tau}\aim + \tfrac{1}{\tau} \uim \\ \vim - v_i \\ a_i \\ -\tfrac{1}{\tau}a_i + \tfrac{1}{\tau} u_i
    \end{bmatrix}, \, i \in S_m \backslash \{1\},
\end{align}
where $d_i = q_{i-1} - q_i - L_i$ is the distance between vehicle $i$ and its preceding vehicle $i-1$, with $q_i$ denoting the position of the vehicle's rear bumper, and $L_i$ its length. Furthermore, $v_i$ and $a_i$ denote the velocity and acceleration of vehicle $i$, respectively, and $S_m \coloneqq \{ i \in \mathbb{N} \mid 1 \leq i \leq m \}$ (i.e., the set of all vehicles in a platoon of length $m \in \mathbb{N}$). The desired acceleration $u_i$ represents the control input of vehicle $i$, and $\tau>0$ is a time constant modeling driveline dynamics.

The control objective for each follower vehicle is to keep a desired inter-vehicle distance $d_{r,i} = r + h v_i, \, i \in S_m \backslash \{ 1 \}$, with time gap $h>0$, and standstill distance $r>0$. The spacing error is defined as $e_i \coloneqq d_i - d_{r,i}$. To ensure the desired spacing while ensuring string-stable vehicle-following behavior,~\cite{ploeg_design_2011} propose a CACC scheme for a homogeneous platoon, formulated as a dynamic controller of the following form: 
\begin{align} 
\label{eq:SD_BaseController} \vspace{-2mm} 
        \mathcal{F} &\coloneqq \left\{ 
            \begin{aligned}
                u_i &= \rho_i, \\
                \dot{\rho}_i &= -\tfrac{1}{h} \rho_i + \tfrac{1}{h}(k_p e_i + k_d \dot{e}_i) + \tfrac{1}{h} \uim,
            \end{aligned} \right.           
\end{align}
where $k_p>0$ and $k_d>0$ are control gains to be designed.

Each vehicle in $S_m$ is equipped with onboard sensors (e.g., LiDAR, cameras, and velocity/acceleration sensors) and wireless V2V communication to receive data from adjacent vehicles. Considering a combination of sensors, we assume to have access to the following sensor data:
\begin{align} \vspace{-4mm}  \label{eq:SD_Sensors}
    \begin{aligned}
        y_{i,1} &\coloneqq d_i + \delta_{i,1}, 
        \ \ \ \ \ \ \ \ y_{i,2} \coloneqq v_i + \delta_{i,2}, \\
        y_{i,3} &\coloneqq a_i + \delta_{i,3}, 
        \ \ \ \ \ \ \ \ y_{i,4} \coloneqq v_{i-1} - v_i + \delta_{i,4}, \\
        y_{i,5} &\coloneqq a_{i-1} + \delta_{i,5}, 
        \ \ \ \ \ y_{i,6} \coloneqq u_{i-1} + \delta_{i,6}.
    \end{aligned}
\end{align} \vspace{0mm} 
Herein, $\delta_{i,j}$ models potential FDI attacks. Furthermore, sensors $y_{i,1}$ and $y_{i,4}$ provide relative distance and velocity information, $y_{i,2}$ and $y_{i,3}$ are onboard measured velocity and acceleration, while $y_{i,5}$ and $y_{i,6}$ represent wirelessly received data from the preceding vehicle. 

To enhance the robustness of the CACC scheme in~\eqref{eq:SD_BaseController} against the FDI attacks in~\eqref{eq:SD_Sensors}, we propose a hybrid control framework that exploits equivalent controller realizations, i.e., distinct realizations producing the same control input~$u_i$ while relying on different sensors. Using the measurements $y_i=[y_{i,1}, \dots, y_{i,6}]^\top$, we define a new controller state $\bar{\rho}_i$ via the linear change of coordinates, see \cite{huisman_optimal_2024}: 
\begin{align} \label{eq:CR_Transformation}
  \bar{\rho}_i &=  \rho_i + \begin{bmatrix} \beta_{i,1} & \beta_{i,2} & \beta_{i,3} & \beta_{i,4} & \beta_{i,5} & 0 \end{bmatrix} = \rho_i + \beta_i y_i.
\end{align}
This transformation yields a class of equivalent controller realizations that produce the same control input $u_i$ while depending on distinct sensor subsets. Applying~\eqref{eq:CR_Transformation} to~\eqref{eq:SD_BaseController} yields:
\begin{subequations}\label{eq:rhoBar} \vspace{0mm}  \begin{align} 
    u_i &= \bar\rho_i - \beta_i y_i, \\
    \dot{\bar{\rho}}_i &= -\left(\tfrac{1}{h} + \tfrac{\beta_{i,3}}{\tau} \right)
 \bar\rho_i + \sum\nolimits_{j=1}^{6} \bar B_{i,j} y_{i,j}, \\
     \bar B_{i,1} &= -\tfrac{\beta_{i,1} \beta_{i,3}}{\tau}-\tfrac{\beta_{i,1}}{h}+\tfrac{k_p}{h},\\
     \bar B_{i,2} &= -\tfrac{\beta_{i,2} \beta_{i,3}}{\tau}-\tfrac{\beta_{i,2}}{h}-k_p,\\
     \bar B_{i,3} &= \beta_{i,4}-\tfrac{\beta_{i,3}^2}{\tau}-\beta_{i,2}+\tfrac{\beta_{i,3}}{\tau}-\tfrac{\beta_{i,3}}{h}-k_d,\\
     \bar B_{i,4} &= \tfrac{k_d}{h}-\tfrac{\beta_{i,4} \beta_{i,3}}{\tau}-\beta_{i,1}-\tfrac{\beta_{i,4}}{h},\\
     \bar B_{i,5} &= \tfrac{\beta_{i,5}}{\tau}-\tfrac{\beta_{i,5} \beta_{i,3}}{\tau}-\beta_{i,4}-\tfrac{\beta_{i,5}}{h},\,\,\,
     \bar B_{i,6} = \tfrac{1}{h}-\tfrac{\beta_{i,5}}{\tau},
 \end{align} \end{subequations}
where each pair ($\beta_{i,j},\bar{B}_{i,j}$) yields another controller realization, which in itself can depend on a different subset of sensors.

By operating several such realizations in parallel, each using a distinct subset of sensors, any inconsistency among their control signals indicates the presence of an FDI attack. This motivates a hybrid control architecture that monitors these realizations and switches to a healthy controller upon detecting an attack. Our contributions are: 1) a method to construct distinct minimal-sensor equivalent controller realizations (Section~\ref{sec:controller_realization}), and 2) a hybrid attack detection and mitigation scheme that switches between these realizations, ensuring a healthy control input is always applied (Section~\ref{sec:Hybrid_Control_Scheme}). Importantly, a dedicated stability analysis of the hybrid closed-loop system is not required, as the proposed scheme ensures that the healthy control input coincides with the nominal input of \eqref{eq:SD_BaseController}, for which stability and the desired closed-loop performance are already established.

%% file: 4.Approach.tex
\section{Controller Realizations} \label{sec:controller_realization} \vspace{-2mm}
\input{4_Approach/4.1_ControllerRealizationsV3}

\section{Hybrid Control Scheme} \vspace{-2mm} \label{sec:Hybrid_Control_Scheme}
\input{4_Approach/4.2_HybridDynamicsV2}

%% file: 4_Approach/4.1_ControllerRealizationsV3.tex
Although $y_i$ comprises six sensor measurements, we first seek to minimize the sensor dependency. Reducing the number of sensors lowers both redundancy and security requirements~(\cite{anand_risk_2022}). This is achieved by finding pairs ($\beta_{i,j},\bar{B}_{i,j}$) equal to zero, where $\beta_{i,j} = 0$ ensures that the sensor is not required in \eqref{eq:CR_Transformation}, and $\bar{B}_{i,j}$ excludes the sensor in \eqref{eq:rhoBar}. For example, excluding $y_{i,1}$ requires
\begin{align} \vspace{-1mm}
    \beta_{i,1} &= 0, \quad 
    \bar{B}_{i,1} = -\tfrac{\beta_{i,1} \beta_{i,3}}{\tau}-\tfrac{\beta_{i,1}}{h}+\tfrac{k_p}{h} = 0,
\end{align}
which has no solution for $h,k_p>0$, indicating that $y_{i,1}$ is always required to realize the controller. In contrast, excluding $y_{i,6}$ is possible by satisfying
\begin{align}  \label{eq:beta5_y5} \vspace{-1mm}
    \bar B_{i,6} &= \tfrac{1}{h}-\tfrac{\beta_{i,5}}{\tau} =0,
\end{align} 
implying $\beta_{i,5} = \tfrac{\tau}{h}$. This implies that excluding both $y_{i,5}$ and $y_{i,6}$ is impossible as this would require $\beta_{i,5} = 0$. Similar results hold for other sensors, but we omit the derivations for brevity.

Following this approach, we obtain two controller realizations $\mathcal{F}_{i,1}$ and $\mathcal{F}_{i,2}$ requiring only three sensors. By selecting $\beta_{i}^1 = \begin{bmatrix} \tfrac{k_d}{h} & -k_d & 0 & 0 & \tfrac{\tau}{h} & 0\end{bmatrix}$ and $\beta_{i}^2 = \begin{bmatrix} \tfrac{k_d}{h} & {-k_d} & 0 & 0 & 0 &, 0\end{bmatrix}$, respectively, we obtain:
\begin{subequations} \label{eq:Ci1Ci2} \begin{align}
    \mathcal F_{i,1}&= \left\{ 
    \begin{aligned} 
        \dot \rho_{i,1} &=  -\tfrac{1}{h}\rho_{i,1} +\left(\tfrac{k_p}{h} -\tfrac{k_d}{h^2}\right)\,y_{i,1} \\ & -\left(k_p-\tfrac{k_d}{h}\right)y_{i,2} -\left(\tfrac{\tau }{h^2}-\tfrac{1}{h}\right)y_{i,5}, \\
        u_{i,1} &= \rho_{i,1} + \tfrac{k_d}{h} y_{i,1} - k_d y_{i,2}+\tfrac{\tau}{h}y_{i,5},
    \end{aligned}
    \right. \\ 
    \mathcal F_{i,2}&= \left\{ 
    \begin{aligned}
        \dot \rho_{i,2} &= -\tfrac{1}{h}\rho_{i,2} + \left(\tfrac{k_p}{h}-\tfrac{k_d}{h^2}\right)y_{i,1} \\
        &- \left(k_p-\tfrac{k_d}{h}\right)y_{i,2} +\tfrac{1}{h}y_{i,6}, \\
        u_{i,2} &= \rho_{i,2} + \tfrac{k_d}{h} y_{i,1} -k_d y_{i,2},
    \end{aligned}
    \right. 
\end{align}\end{subequations}
with the new controller states $\rho_{i,1}$ and $\rho_{i,2}$. Since these realizations are independent of the vehicle index $i$, there is no longer a need to distinguish between vehicles $i$ and $i-1$. Therefore, the subscript $i$ is omitted for clarity hereafter.  
\begin{assumption} \label{ass:one_attack}
The attacker may switch between sensors arbitrarily, but may compromise only one sensor at any given time.
\end{assumption}
Since both $\mathcal{F}_{1}$ and $\mathcal{F}_{2}$ rely on $y_1$ and $y_2$, we assume that the vehicle is equipped with three independent measurements of $y_{1|k}$ and $y_{2|k}$, where $y_{j|k}$ denotes the $k$th physical measurement of signal $y_j$. For instance, $y_{1|1}$ is obtained via LiDAR, while $y_{1|2}$ and $y_{1|3}$ are obtained via different cameras. The availability of three physical copies allows their true values to be recovered by majority voting under Assumption~\ref{ass:one_attack}. For example, if $y_{1|3}$ is compromised, then $y_{1|1}=y_{1|2}\neq y_{1|3}$, and the correct value of $y_1$ is obtained from the majority. 

For $y_5$ and $y_6$, direct majority voting is not possible as the signals are distinct. This implies that achieving resilience through standard voting would require either tripling $y_5$ or $y_6$. To avoid this hardware redundancy, we instead exploit the fact that the controller realizations in~\eqref{eq:Ci1Ci2} produce the same nominal input in the absence of attacks. Specifically, we construct four controller realizations using two measurements of $y_{5|k}$ and $y_{6|k}$, yielding the sensor set $\mathcal{Y}\!=\!\{y_1, y_2, y_{5|k}, y_{6|k}\}$ for $k\in\{1,2\}$, and define
\begin{subequations} \label{eq:C1zC2z} \begin{align}
    \mathcal F_{1|k}&= \left\{ 
    \begin{aligned} 
        \dot \rho_{1|k} &=  -\tfrac{1}{h}\rho_{1|k} +\left(\tfrac{k_p}{h} -\tfrac{k_d}{h^2}\right)\,y_{1} \\ & -\left(k_p-\tfrac{k_d}{h}\right)y_{2} -\left(\tfrac{\tau }{h^2}-\tfrac{1}{h}\right) y_{5|k}, \\
        u_{1|k} &= \rho_{1|k} + \tfrac{k_d}{h} y_{1} - k_d y_{2}+\tfrac{\tau}{h} y_{5|k},
    \end{aligned}
    \right. \end{align}\begin{align}
    \mathcal F_{2|k}&= \left\{ 
    \begin{aligned}
        \dot \rho_{2|k} &= -\tfrac{1}{h}\rho_{2|k} + \left(\tfrac{k_p}{h}-\tfrac{k_d}{h^2}\right)y_{1} \\
        &- \left(k_p-\tfrac{k_d}{h}\right)y_{2} +\tfrac{1}{h} y_{6|k}, \\
        u_{2|k} &= \rho_{2|k} + \tfrac{k_d}{h} y_{1} -k_d y_{2},
    \end{aligned}
    \right. 
\end{align}\end{subequations}
where $\rho_{1|k}$ and $\rho_{2|k}$ denote the new controller states and $k$ denoting which sensor is used. For example, $\mathcal F_{1|1}$ utilizes $y_{5|1}$, and $\mathcal F_{1|2}$ utilizes $y_{5|2}$. Since these controller realizations are equivalent in the absence of an attack, the following holds
\begin{subequations} \vspace{0mm} \label{eq:RealizationProperties}
    \begin{align} 
        u_{1|1} &= u_{1|2} = u_{2|1} = u_{2|2}, \\
        \implies \rho_{1|1} + \tfrac{\tau}{h} y_{5|1} &= \rho_{1|2} + \tfrac{\tau}{h}y_{5|2} =  \rho_{2|1} = \rho_{2|2}. 
    \end{align}
\end{subequations}
Violations of the equalities in~\eqref{eq:RealizationProperties} indicate an FDI attack, i.e, $\delta_{4+j|k} \neq 0, \, j,k \in \{1,2\}$. The next section presents a hybrid control scheme that leverages the controller realizations~\eqref{eq:C1zC2z} and its properties~\eqref{eq:RealizationProperties} to identify the compromised sensors and recover a healthy control input.

%% file: 4_Approach/4.2_HybridDynamicsV2.tex
This section presents the proposed hybrid control scheme to detect FDI attacks on sensor measurements $y_{4+j|k}$ for $j,k\in \{1,2\}$, and determine a healthy control input $u^*$ that preserves the nominal control performance of \eqref{eq:SD_BaseController}. The control scheme is modeled as a hybrid automaton, partitioned into a continuous state $\rho$ and a discrete mode $q$. We adopt the following hybrid systems formulation of \cite{goebel_hybrid_2012}, yielding:
\begin{align} \label{eq:HybridSystem_rho}
    \left\{ \begin{aligned}
    \dot{\rho} &= f(\rho, \mathcal{Y}), &&\text{ if }( \rho, q, \mathcal{Y})\in \mathcal C, \\
        (\rho,q)^+ &= g(\rho, q, \mathcal{Y}), &&\text{ if }(\rho, q, \mathcal{Y})\in \mathcal D, \\
        u^* &= h(\rho, q, \mathcal{Y}),
    \end{aligned} \right.
\end{align}
where $\rho = [\rho_{1|1}, \rho_{1|2}, \rho_{2|1}, \rho_{2|2}]^\top$ denotes the stacked controller states, $\dot{\rho}$ denotes their rate of change, $\rho^+$ and $q^+$ denote the updated state and mode after an instantaneous jump, respectively, and $\mathcal{Y}$ denotes the sensor measurements. The functions $f(\rho, \mathcal{Y})$, $g(\rho, q, \mathcal{Y})$, and $h(\rho, q, \mathcal{Y})$ define the flow map, jump map, and output map, respectively, with the corresponding flow set $\mathcal{C}$ and jump sets $\mathcal{D}$. Definition~\ref{DefHybridSystem} specifies the solution notion adopted for the hybrid system.

To account for the different attack scenarios, we define a set of discrete system modes $Q~=~\{q_0, q_{1|1}, q_{1|2}, q_{2|1}, q_{2|2}\}$, each corresponding to a different attack scenario where: 
\begin{subequations} \label{eq:modes} \begin{align}
    q_{0} &: \delta_{4+j|k} = 0, \text{ i.e., healthy}, \\
    q_{j|k}&: \delta_{4+j|k} \neq 0, \text{ i.e., attack on }\mathcal F_{j|k}, \, j,k\in \{1,2\}.
\end{align} \end{subequations}
Each mode governs which controller outputs are trusted and how to mitigate the FDI attack. For example, in $q_0$, all controllers operate nominally. In contrast, in $q_{1|1}$, an attack on $y_{5|1}$ implies that $u_{1|1}$ is unreliable and must therefore be excluded to determine a healthy control input $u^*$.

In the following subsections, for each mode $q\in Q$, the different flow and jump conditions are defined. The first subsection describes the flow behavior and the subsequent subsections elaborate on three distinct jumping behaviors,  namely; i) Jumps to/from the healthy mode $q_0$, ii) jumps between $q_{j|k}$ and $q_{3-j|l}$ for $j,k,l \in \{1,2\}$, e.g., switching from $y_{5|1}$ to $y_{6|1}$, and iii) jumps between $q_{j|k}$ and $q_{j|3-k}$ for $j,k \in \{1,2\}$, e.g., switching from $y_{5|1}$ to $y_{5|2}$. 
  
\subsection{Flow Design} \vspace{-1mm}
The flow map $f$ is independent of the mode $q$, and yields
\begin{align}
    f(\rho, \mathcal{Y}) = \begin{bmatrix}
        \dot{\rho}_{1|1} & \dot{\rho}_{1|2} & \dot{\rho}_{2|1} & \dot{\rho}_{2|2}
    \end{bmatrix}^\top,
\end{align}
which is identical for all $q\in Q$, with the corresponding dynamics described by the right-hand side of~\eqref{eq:C1zC2z}. 

The flow set $\mathcal C$ is determined based on the controller realizations' properties in~\eqref{eq:RealizationProperties}. These equalities hold in the absence of an attack but are violated if $\delta_{4+j|k} \neq 0, \, j,k \in \{1,2\}$. For example, if $\delta_{5|1} \neq 0$, we observe:
\begin{subequations} \label{eq:Cq1}\begin{align}
    y_{5|1} &\neq y_{5|2}, \quad y_{6|1} = y_{6|2},\\ 
    u_{1|1} &\neq u_{1|2} = u_{2|1} = u_{2|2}.
\end{align}\end{subequations}
These conditions persist until the attack stops or switches to a different sensor. During this attack, the conditions in~\eqref{eq:Cq1} are satisfied, and the hybrid controller flows in mode $q_{1|1}$. While flowing in $q_{1|1}$, it is sufficient to only monitor the healthy controllers, i.e., $u_{1|2} = u_{2|1} = u_{2|2}$. If there is a deviation in the healthy controllers, another sensor may be compromised, prompting a mode jump. Applying this logic across all $q \in Q$, we define the flow sets as:
\begin{subequations}
\begin{align}
    \hspace{-4mm} \mathcal C_0 &\coloneqq \{(\rho,\mathcal Y)\mid u_{1|1}=u_{1|2}=u_{2|1}=u_{2|2} \}, \label{eq:flowset_0}
\end{align} \vspace{-4mm} \begin{multline}
    \mathcal C_{j|k} \coloneqq\{(\rho, \mathcal Y)\mid y_{4+j|1}\neq y_{4+j|2}, \\
     u_{j|3-k}=u_{3-j|1}=u_{3-j|2} \} \,\, j,k\in\{1,2\}, \label{eq:flowset_jk}
    \end{multline}
\end{subequations}
leading to the overall flow set for \eqref{eq:HybridSystem_rho}:
\begin{align} \label{eq:Flowset}
    \mathcal{C} \coloneqq \mathcal C_0 \bigcup_{j,k \in \{1,2\}} \mathcal C_{j|k}.
\end{align} 

\subsection{Jump Design to and from Healthy Mode} 
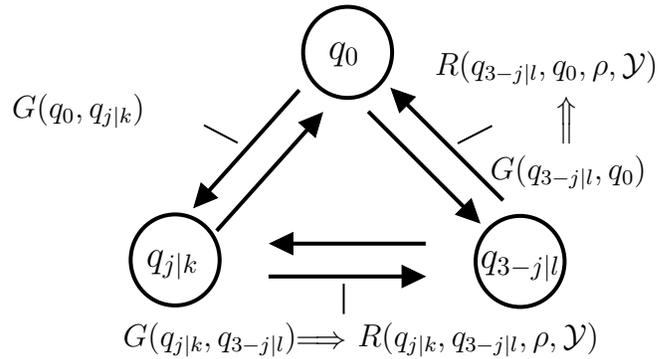
\begin{figure}[bt]
    \centering
    \begin{adjustbox}{width=0.41\textwidth}
        \input{00_Figures/ModesScheme_general} 
    \end{adjustbox}\vspace{-3mm}
    \caption{Jumps between modes $q_0, q_{j|k}$, and $q_{3-j|l}, \, j,k,l\in\{1,2\}$, where the system either jumps from or to healthy mode, or transitions between attack modes with different controller realizations $\mathcal{F}_{j|k}$ and $\mathcal{F}_{3-j|l}$.}
    \label{fig:ModesScheme_external} 
    \vspace{0mm}
\end{figure}

\vspace{-1mm}
Jumps from or to the healthy mode $q_0$ occur when the attacker initiates or stops an FDI attack, respectively. These jumps are governed by so-called \emph{guards}, following the notation in \cite[Section~1.4.2]{goebel_hybrid_2012}, where a jump from mode $q$ to $q'$ is triggered when the system state belongs to the guard set $G(q,q')$. As depicted in Fig.~\ref{fig:ModesScheme_external}, for $j,k \in \{1,2\}$, the jump from $q_0$ to $q_{j|k}$ is governed by the guard $G(q_0, q_{j|k})$, while the jump from $q_{j|k}$ back to $q_0$ is governed by $G(q_{j|k}, q_0)$.

When the hybrid controller flows in $q_0$, i.e., when \eqref{eq:flowset_0} is satisfied, a jump to another mode $q_{j|k}$, $j,k \in \{1,2\}$, is triggered if the flow conditions are violated. For instance, when $\delta_{5|1} \neq 0$, the control inputs no longer match, i.e.,
 \begin{align}
     u_{1|1} \neq u_{1|2} = u_{2|1} = u_{2|2}.
 \end{align}
Similar deviations occur for any $\delta_{4+j|k} \neq 0$. These conditions define the guard sets for jumps from $q_0$:
\begin{multline}
    G(q_{0}, q_{j|k}) = \{(\rho, \mathcal Y\} \mid u_{j|k}  \!\neq\!  u_{3-j|k} \!=\! u_{1|3-k}  \!=\!  u_{2|3-k} \},\\ \quad j,k\in\{1,2\}
\end{multline}

Returning from an attack mode $q_{j|k}$ to $q_0$ requires detecting whether the FDI attack has stopped. Since $\delta_{4+j|k}$ affects the controller dynamics $\dot{\rho}_{j|k}$, the state $\rho_{j|k}$ is invalid even if $\delta_{4+j|k} = 0$. Thus, control inputs remain mismatched:
\begin{align} \label{eq:ControlMismatchq0}
u_{j|k} \neq u_{3-j|k}=u_{1|3-k}=u_{2|3-k}, 
\end{align}
despite the sensor equality $y_{4+j|1} = y_{4+j|2}$. Therefore, detecting whether the attack has stopped relies on both sensor measurements and control input equivalences. The corresponding guard sets are defined as:
\begin{multline} 
   G(q_{j|k}, q_{0}) \! \coloneqq \! \{(\rho, \mathcal Y\} \mid y_{4+j|1} \! = \! y_{4+j|2}, \\ u_{j|3-k} \! = \! u_{3-j|1} \! = \! u_{3-j|2}\}, \quad j,k\in \{1,2\}.
 \end{multline}\vspace{-5mm}

If the guard $G(q_{j|k}, q_{0})$ is activated, the control mismatch in \eqref{eq:ControlMismatchq0} persists. However, in mode $q_{j|k}$, the controllers $\mathcal F_{3-j|k}$, $\mathcal F_{1|3-k}$, and $\mathcal F_{2|3-k}$ remain valid. To prevent any lingering effects from the attack upon returning to $q_0$, we apply a reset based on \eqref{eq:RealizationProperties}, defined as
\begin{align}
     R(q_{j|k}, q_{0},  \rho, \mathcal Y): \quad \rho_{j|k}^+ \coloneqq \rho_{j|3-k}, \quad j,k\in\{1,2\}.
\end{align}
Here, the previously compromised controller state  $\rho_{j|k}$ is reset using a valid state $\rho_{j|3-k}$. This ensures that, upon returning to $q_0$, the control inputs satisfy $u_{1|1} = u_{1|2} = u_{2|1} = u_{2|2}$, i.e., satisfy \eqref{eq:flowset_0}.

The resulting jump set $\mathcal D_0$, corresponding to mode $q_0$, is then defined as
\begin{align} \begin{split}
    \mathcal D_0 \coloneqq& \bigcup_{j,k \in \{1,2\}} G(q_{0}, q_{j|k})    \cup G(q_{j|k}, q_{0}).
\end{split} \end{align}

\subsection{Jump Design Unhealthy Modes} \vspace{-1mm}
An attacker may alter the FDI attack by switching to a different sensor. The hybrid controller consists of two realizations, $\mathcal F_{j|k}$ and $\mathcal F_{3-j|l}$, each with two variants $\mathcal F_{j|k}$ and $\mathcal F_{j|k-3}$, for $j,k,l \in \{1,2\}$. This results in two distinct types of jumps between controller realizations.

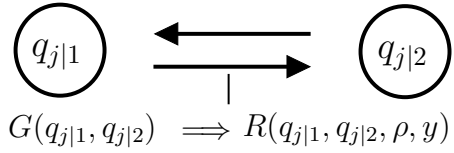
\begin{figure}[tb]
    \centering
    \begin{adjustbox}{width=0.325\textwidth}
    \input{00_Figures/modeScheme_internal}
    \end{adjustbox} \vspace{-3mm}
    \caption{Switching between modes $q_{j|1}$ and $q_{j|2}$ for $j\in\{1,2\}$, indicating the attacker switches its FDI attack between $\mathcal F_{j|1}$ and $\mathcal F_{j|2}$.}
    \label{fig:ModesScheme_internal}
    \vspace{0mm}
\end{figure}

The first type, depicted in Fig.~\ref{fig:ModesScheme_external}, involves jumps between $q_{j|k}$ and $q_{3-j|l}$ for $j,k,l\in\{1,2\}$, triggered by guards $G(q_{j|k}, q_{3-j|l})$, with reset $R(q_{j|k}, q_{3-j|l}, \rho, \mathcal Y)$. Such jumps occur when the attacker switches between the sensors $y_{5|k}$ and $y_{6|l}$. The second type, depicted in Fig.~\ref{fig:ModesScheme_internal}, concerns jumps between $q_{j|k}$ and $q_{j|3-k}$, governed by guards $G(q_{j|k}, q_{j|3-k})$, with reset $R(q_{j|k}, q_{j|3-k}, \rho, \mathcal Y)$. These occur when an attacker switches between sensors $y_{4+j|k}$ and $y_{4+j|3-k}$.

The distinction between these two jump types is based on how the flow conditions in \eqref{eq:flowset_jk} are violated when the attacker switches sensors. While the system flows in $q_{j|k}$, i.e., when \eqref{eq:flowset_jk} holds, a jump to $q_{3-j|l}$ must be initiated when
\begin{subequations} \label{eq:JumpCondClass} \begin{align} 
     y_{4+j|1} &= y_{4+j|2}, \\
     u_{j|3-k} &= u_{3-j|3-l} \neq u_{3-j|l}, \quad j,k,l \in \{1,2\},
\end{align}\end{subequations}
is observed. For example, if the attacker switches from $y_{5|1}$ to $y_{6|2}$, we observe that $y_{5|1} = y_{5|2}$ and $u_{1|2} = u_{2|1} \neq u_{2|2}$. 

On the other hand, when the system flows in $q_{j|k}$, a jump to $q_{3-j|l}$ must be initiated when we observe:
\begin{subequations} \label{eq:JumpCondVariant} \begin{align}
     y_{4+j|1} &\neq y_{4+j|2}, \\
     u_{j|3-k} &\neq u_{3-j|1} = u_{3-j|2}, \quad j,k,l \in \{1,2\}.
\end{align} \end{subequations}
For instance, when the attacker switches from $y_{5|1}$ to $y_{5|2}$, and observe that $y_{5|1} \neq y_{5|2}$ and $u_{1|2} \neq u_{2|1} = u_{2|2}$.

Conditions \eqref{eq:JumpCondClass} and \eqref{eq:JumpCondVariant} represent distinct conditions under Assumption~\ref{ass:one_attack}. Consequently, it is always possible to identify the compromised sensor and determine a healthy control input $u^*$, even when the attacker switches between sensors. Depending on how the flow set \eqref{eq:flowset_jk} is violated, a jump is triggered according to the following guards:
\begin{subequations} \vspace{0mm}
\begin{multline}
    G(q_{j|k},q_{3-j|l})= \{(\rho,\mathcal Y) \mid y_{4+j|1} = y_{4+j|2}, \\
      u_{j|3-k}=u_{3-j|3-l}\neq u_{3-j|l} \}, \quad j,k,l\in\{1,2\}, 
\end{multline} \vspace{-4mm} \begin{multline}
    G(q_{j|k},q_{j|3-k}) = \{(\rho,\mathcal Y) \mid y_{4+j|1}\neq y_{4+j|2}, \\
     u_{j|3-k}\neq u_{3-j|1}=u_{3-j|2} \}, \, \, \, \quad \quad \quad j,k\in\{1,2\}. 
\end{multline}  
\end{subequations} \vspace{-5mm}

Upon a jump, the invalid internal controller state must be reset. The reset depends on which guard is triggered. When the attacker switches between $\mathcal{F}_{j|k}$ and $\mathcal{F}_{3-j|l}$ (i.e., between $y_{5|k}$ and $y_{6|l}$), a healthy controller realization $\mathcal{F}_{j|3-k}$ is always available to reset the invalid controller state, similar to the transition to $q_0$. However, when the attacker switches from $y_{6|k}$ to $y_{6|3-k}$, the condition 
\begin{align}
    u_{1|1} = u_{1|2} \neq u_{2|3-k}
\end{align} 
occurs because $\rho_{2|k-3}$ is invalid due to $\delta_{6|k-3}$ only affecting $\dot{\rho}_{2|k}$ in \eqref{eq:C1zC2z}. Hence, $\rho_{2|1}$ cannot be reset using $\rho_{2|2}$. Instead, via \eqref{eq:RealizationProperties}, the reset is defined as
\begin{align}
    \rho_{2|k}^+ \coloneqq \rho_{1|l} + \tfrac{\tau}{h}y_{5|l}, \quad k,l \in \{1,2\}.
\end{align}
Applying this approach for all $q \in Q \setminus \{q_0\}$, the resets for $j,k \in\{1,2\}$ are:
\begin{subequations}\begin{align}
      & \hspace{-1.5mm} R(q_{j|k},q_{3-j|l},\rho, \mathcal Y): \quad \rho_{j|k}^+=\rho_{j|3-k}, \\
      & \hspace{-1.5mm} R(q_{j|k}, q_{j|3-k}, \rho, \mathcal Y):  \quad \rho_{j|k}^+=\rho_{3-j|k}+(-1)^j\tfrac{\tau}{h} y_{5|l}.
\end{align}\end{subequations} \vspace{-5mm}

The jump sets $D_{j|k}$, governing all jumps and resets between unhealthy modes, are then defined as  
\begin{align} \begin{split}
   \hspace{-2.5mm} \mathcal D_{j|k} \coloneqq& \hspace{-3mm} \bigcup_{j,k,l\in \{1,2\}} G(q_{j|k},q_{3-j|l}) \cup G(q_{j|k},q_{j|3-k}).
\end{split} \end{align}
The resulting jump set $\mathcal D$ for \eqref{eq:HybridSystem_rho} can be defined as the set of all possible jumps between modes $q \in Q$ that the guards and the corresponding resets trigger:
\begin{align} \label{eq:JumpSet}  \begin{split}
    \mathcal D \coloneqq & \mathcal D_0 \bigcup_{j,k \in \{1,2\}} \mathcal D_{j|k}.
\end{split} \end{align}

Finally, the control input $u^*$ is selected based on the current mode $q \in Q$. In mode $q_0$ (where \eqref{eq:flowset_0} holds), any of the inputs $u_{1|1}, u_{1|2}, u_{2|1}$, or $u_{2|2}$ from \eqref{eq:C1zC2z} may be chosen. For mode $q_{j|k}$ (where \eqref{eq:flowset_jk} holds), $u^*$ is selected from $\{u_{j|3-k}, u_{3-j|1}, u_{3-j|2}\}$. Thus, for $j,k \in \{1,2\}$:
\begin{align}
    u^* \in \{u_{1|1}, u_{1|2}, u_{2|1}, u_{2|2}\} \setminus \{u_{j|k}\}, \quad \text{if } q = q_{j|k}.
\end{align}
Under Assumption~\ref{ass:one_attack} and the instantaneous switching, the controller input $u^*$ coincides with the nominal input~\eqref{eq:SD_BaseController} at all times. Consequently, the closed-loop dynamics are unchanged, and the nominal stability properties directly apply. 

Furthermore, by combining the majority voting for $\{y_1, y_2\}$ with the hybrid detection scheme for $\{y_{5|k}, y_{6|k}\}$, Assumption~\ref{ass:one_attack} is not limited to one sensor. Instead, the framework remains resilient to simultaneous attacks on a subset of sensors, allowing for one compromised measurement of $y_1$, one of $y_2$, and one from the set $\{y_{5|1}, y_{5|2}, y_{6|1}, y_{6|2}\}$. The next section illustrates the proposed scheme through a simulation study.

%% file: 00_Figures/ModesScheme_general.tex
\tikzset{every picture/.style={line width=0.75pt}} 

\begin{tikzpicture}[x=0.75pt,y=0.75pt,yscale=-1,xscale=1]

\draw  [line width=1.5]  (310.04,114.16) .. controls (310.04,101.47) and (320.09,91.18) .. (332.48,91.18) .. controls (344.87,91.18) and (354.91,101.47) .. (354.91,114.16) .. controls (354.91,126.85) and (344.87,137.14) .. (332.48,137.14) .. controls (320.09,137.14) and (310.04,126.85) .. (310.04,114.16) -- cycle ;
\draw  [line width=1.5]  (223.05,218.93) .. controls (223.05,206.24) and (233.1,195.96) .. (245.49,195.96) .. controls (257.88,195.96) and (267.92,206.24) .. (267.92,218.93) .. controls (267.92,231.62) and (257.88,241.91) .. (245.49,241.91) .. controls (233.1,241.91) and (223.05,231.62) .. (223.05,218.93) -- cycle ;
\draw  [line width=1.5]  (397.03,219.69) .. controls (397.03,207) and (407.08,196.71) .. (419.47,196.71) .. controls (431.86,196.71) and (441.9,207) .. (441.9,219.69) .. controls (441.9,232.38) and (431.86,242.67) .. (419.47,242.67) .. controls (407.08,242.67) and (397.03,232.38) .. (397.03,219.69) -- cycle ;
\draw [line width=1.5]    (371.86,210.7) -- (297.27,210.7) ;
\draw [shift={(293.27,210.7)}, rotate = 360] [fill={rgb, 255:red, 0; green, 0; blue, 0 }  ][line width=0.08]  [draw opacity=0] (11.61,-5.58) -- (0,0) -- (11.61,5.58) -- cycle    ;
\draw [line width=1.5]    (292.59,227.28) -- (367.54,227.28) ;
\draw [shift={(371.54,227.28)}, rotate = 180] [fill={rgb, 255:red, 0; green, 0; blue, 0 }  ][line width=0.08]  [draw opacity=0] (11.61,-5.58) -- (0,0) -- (11.61,5.58) -- cycle    ;
\draw [line width=1.5]    (342.87,144.01) -- (396.27,197.41) ;
\draw [shift={(399.1,200.24)}, rotate = 225] [fill={rgb, 255:red, 0; green, 0; blue, 0 }  ][line width=0.08]  [draw opacity=0] (11.61,-5.58) -- (0,0) -- (11.61,5.58) -- cycle    ;
\draw [line width=1.5]    (410.89,188.87) -- (357.23,135.21) ;
\draw [shift={(354.41,132.38)}, rotate = 45] [fill={rgb, 255:red, 0; green, 0; blue, 0 }  ][line width=0.08]  [draw opacity=0] (11.61,-5.58) -- (0,0) -- (11.61,5.58) -- cycle    ;
\draw [line width=1.5]    (308.09,133.2) -- (257.62,189.52) ;
\draw [shift={(254.95,192.5)}, rotate = 311.87] [fill={rgb, 255:red, 0; green, 0; blue, 0 }  ][line width=0.08]  [draw opacity=0] (11.61,-5.58) -- (0,0) -- (11.61,5.58) -- cycle    ;
\draw [line width=1.5]    (266.41,204.22) -- (317.13,147.63) ;
\draw [shift={(319.8,144.65)}, rotate = 131.87] [fill={rgb, 255:red, 0; green, 0; blue, 0 }  ][line width=0.08]  [draw opacity=0] (11.61,-5.58) -- (0,0) -- (11.61,5.58) -- cycle    ;
\draw    (387.98,160.12) -- (405.33,151.25) ;
\draw    (260,150.92) -- (277,159.83) ;
\draw    (330.5,230.92) -- (330.5,246.25) ;

\draw (324,107.95) node [anchor=north west][inner sep=0.75pt]  [font=\fontsize{14pt}{15pt}\selectfont]  {$q_{0}$};
\draw (232,212.36) node [anchor=north west][inner sep=0.75pt]  [font=\fontsize{13pt}{15pt}\selectfont]  {${\textstyle q_{j|k}}$};
\draw (400,213.31) node [anchor=north west][inner sep=0.75pt]  [font=\fontsize{13pt}{15pt}\selectfont]  {$q_{3-j|l}$};
\draw (403,164.71) node [anchor=north west][inner sep=0.75pt]  [font=\large]  {$G( q_{3-j|l} ,q_{0})$};
\draw (375,111.94) node [anchor=north west][inner sep=0.75pt]  [font=\large]  {$R( q_{3-j|l} ,q_{0} ,\rho ,\mathcal{Y})$};
\draw (439.47,162.69) node [anchor=north west][inner sep=0.75pt]  [font=\large,rotate=-270]  {$\Longrightarrow $};
\draw (161.81,133.57) node [anchor=north west][inner sep=0.75pt]  [font=\large]  {$G( q_{0} ,q_{j|k})$};
\draw (217.81,249.07) node [anchor=north west][inner sep=0.75pt]  [font=\large]  {$G( q_{j|k} ,q_{3-j|l})$};
\draw (336.1,248.94) node [anchor=north west][inner sep=0.75pt]  [font=\large]  {$R( q_{j|k} ,q_{3-j|l} ,\rho ,\mathcal{Y})$};
\draw (305,256.4) node [anchor=north west][inner sep=0.75pt]  [font=\large]  {$\Longrightarrow $};

\end{tikzpicture}

%% file: 00_Figures/modeScheme_internal.tex
\tikzset{every picture/.style={line width=0.75pt}} 

\begin{tikzpicture}[x=0.75pt,y=0.75pt,yscale=-1,xscale=1]

\draw  [line width=1.5]  (232.05,143.93) .. controls (232.05,131.24) and (242.1,120.96) .. (254.49,120.96) .. controls (266.88,120.96) and (276.92,131.24) .. (276.92,143.93) .. controls (276.92,156.62) and (266.88,166.91) .. (254.49,166.91) .. controls (242.1,166.91) and (232.05,156.62) .. (232.05,143.93) -- cycle ;
\draw  [line width=1.5]  (406.03,144.69) .. controls (406.03,132) and (416.08,121.71) .. (428.47,121.71) .. controls (440.86,121.71) and (450.9,132) .. (450.9,144.69) .. controls (450.9,157.38) and (440.86,167.67) .. (428.47,167.67) .. controls (416.08,167.67) and (406.03,157.38) .. (406.03,144.69) -- cycle ;
\draw [line width=1.5]    (380.86,135.7) -- (306.27,135.7) ;
\draw [shift={(302.27,135.7)}, rotate = 360] [fill={rgb, 255:red, 0; green, 0; blue, 0 }  ][line width=0.08]  [draw opacity=0] (11.61,-5.58) -- (0,0) -- (11.61,5.58) -- cycle    ;
\draw [line width=1.5]    (301.59,152.28) -- (376.54,152.28) ;
\draw [shift={(380.54,152.28)}, rotate = 180] [fill={rgb, 255:red, 0; green, 0; blue, 0 }  ][line width=0.08]  [draw opacity=0] (11.61,-5.58) -- (0,0) -- (11.61,5.58) -- cycle    ;
\draw    (339.5,155.92) -- (339.5,171.25) ;

\draw (238.5,137.36) node [anchor=north west][inner sep=0.75pt]  [font=\Large]  {${\textstyle q_{j|1}}$};
\draw (413.01,138.31) node [anchor=north west][inner sep=0.75pt]  [font=\Large]  {$q_{j|2}$};
\draw (226.81,174.07) node [anchor=north west][inner sep=0.75pt]  [font=\large]  {$G( q_{j|1} ,q_{j|2})$};
\draw (345.1,173.94) node [anchor=north west][inner sep=0.75pt]  [font=\large]  {$R( q_{j|1} ,q_{j|2} ,\rho , \mathcal{Y})$};
\draw (314,181.4) node [anchor=north west][inner sep=0.75pt]  [font=\large]  {$\Longrightarrow $};

\end{tikzpicture}

%% file: 5.Results.tex
\section{Case Study Results} \label{sec:Results} \vspace{-2mm}
\input{5.Results/5.1_CaseStudyV3}

%% file: 5.Results/5.1_CaseStudyV3.tex
This section presents the implementation of a hybrid controller~\eqref{eq:HybridSystem_rho} incorporating the controller realizations~\eqref{eq:Ci1Ci2}, and flow and jump sets~\eqref{eq:Flowset} and~\eqref{eq:JumpSet}, respectively. A two-vehicle platoon is considered, consisting of a leader (vehicle 1) and a follower (vehicle 2), modeled using the platooning dynamics in~\eqref{eq:SD_PlatoonDynamics}. The controller parameters are adopted from \cite{ploeg_design_2011}, with an inter-vehicle distance of $r = 3\,$m, driveline dynamics constant $\tau = 0.1\,$s, time headway constant $h = 0.5\,$s, and controller gains $k_p = 0.2\,\text{s}^{-2}$, $k_d = 0.7\,\text{s}^{-1}$.

As a case study, we seek to detect and mitigate the FDI attacks in \eqref{eq:SD_Sensors}, where the attacker can switch between the different sensors without restrictions. Therefore, we consider the following attack sequence for $j,k \! \in \! \{ 1,2\}$:
\begin{align} \label{eq:AttSequence}
    \begin{cases}
        \delta_{5|1} = \cos(25 t),  & 0.5 \leq t < 2.5, \\
        \delta_{6|1} = e^{0.5t},   & 2.5 \leq t < 4.5, \\
        \delta_{6|2} = 2\cos(10t), & 4.5 \leq t < 6.5, \\
        \delta_{5|2} = \cos(t),   & 6.5 \leq t < 8.5, \\
        \delta_{5|1} = e^{0.25t} + \sin(10t), & 8.5 \leq t < 10.0, \\
        \delta_{4+j|k} = 0,  &  \text{otherwise} .
    \end{cases}
\end{align}
This sequence is arbitrarily chosen to trigger all guards and include every possible jump between modes $q \in Q$ within a single simulation, allowing the behavior of the hybrid controller to be observed under diverse attack patterns. The simulation is available at \url{https://github.com/mischa907/HybridAutomaton_Huisman}.

\begin{figure}[bt]\centering
    \vspace{-3mm}
    \includegraphics[width=\linewidth]{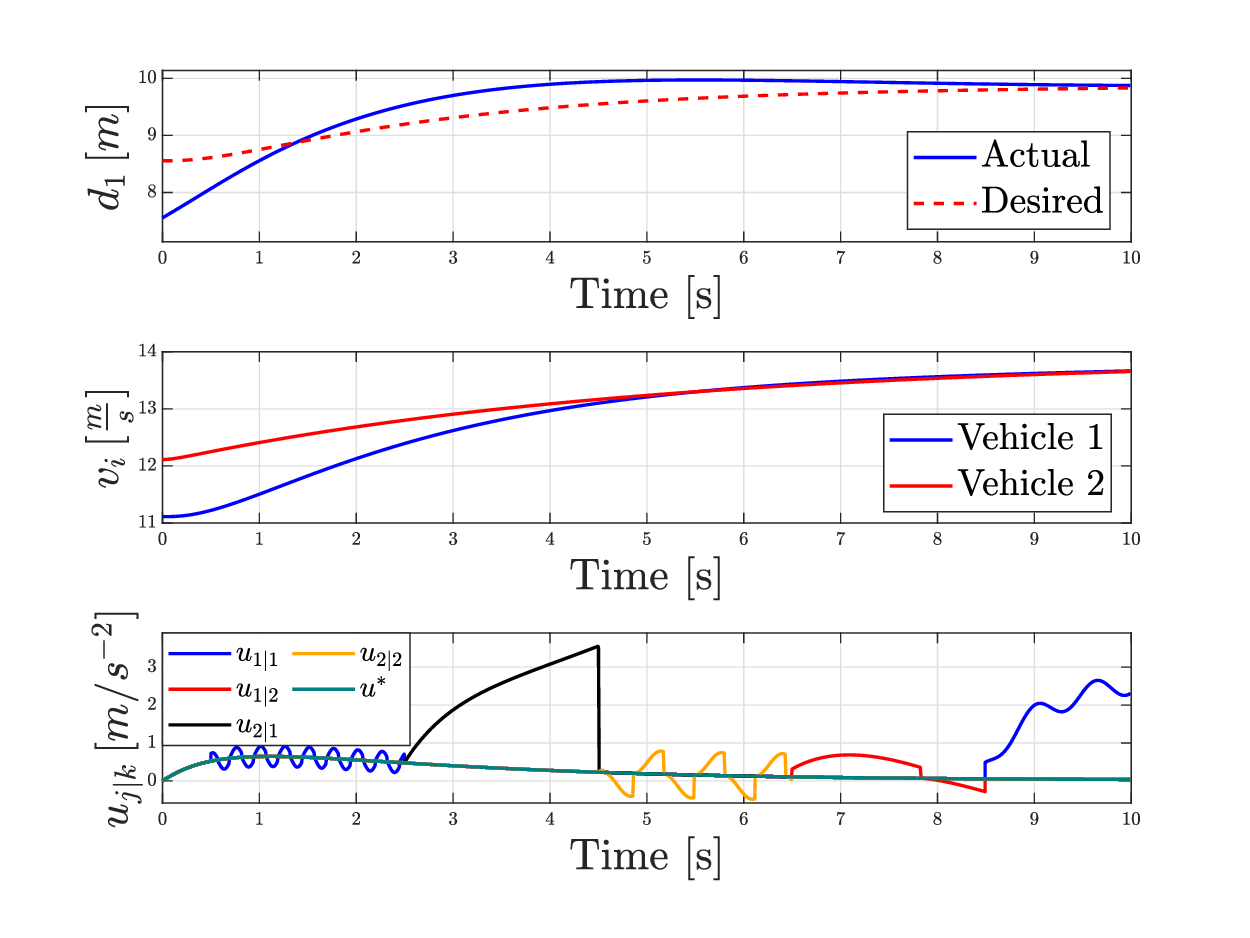}
    \vspace{-7.5mm}
    \caption{Resulting inter-vehicle distance $d_1$, velocities $v_1$ and $v_2$, and controller inputs $u_{j|k}$ for $j,k \in \{1,2\}$ during the FDI sequence \eqref{eq:AttSequence}.}
    \centering
    \label{fig:transient_VehBehavior}
    \vspace{-2mm}
\end{figure}

The scenario begins with transient platoon behavior, i.e.,  $d_1 \neq d_{r,1}$ and $v_1 \neq v_2$, while the leader aims to maintain a velocity of $50\ \mathrm{km/h}$. Fig.~\ref{fig:transient_VehBehavior} illustrates the resulting vehicle behavior in terms of inter-vehicle distance $d_1$, velocity $v_1,v_2$, and the different controller inputs $u_{j|k}$ and $u^*$. Despite the FDI attack, the hybrid controller ensures that the platoon remains unaffected and converges to the desired trajectory. These results show that the FDI does not influence $u^*$, as it does not follow any control inputs deviating due to $\delta_{j|k} \neq 0$ for $j,k \in \{1,2\}$. Additionally, the results highlight the controller resets, as no lingering effects persist in the previously compromised controller when the attacker switches.

\begin{figure}[bt]\centering
    \includegraphics[width=\linewidth]{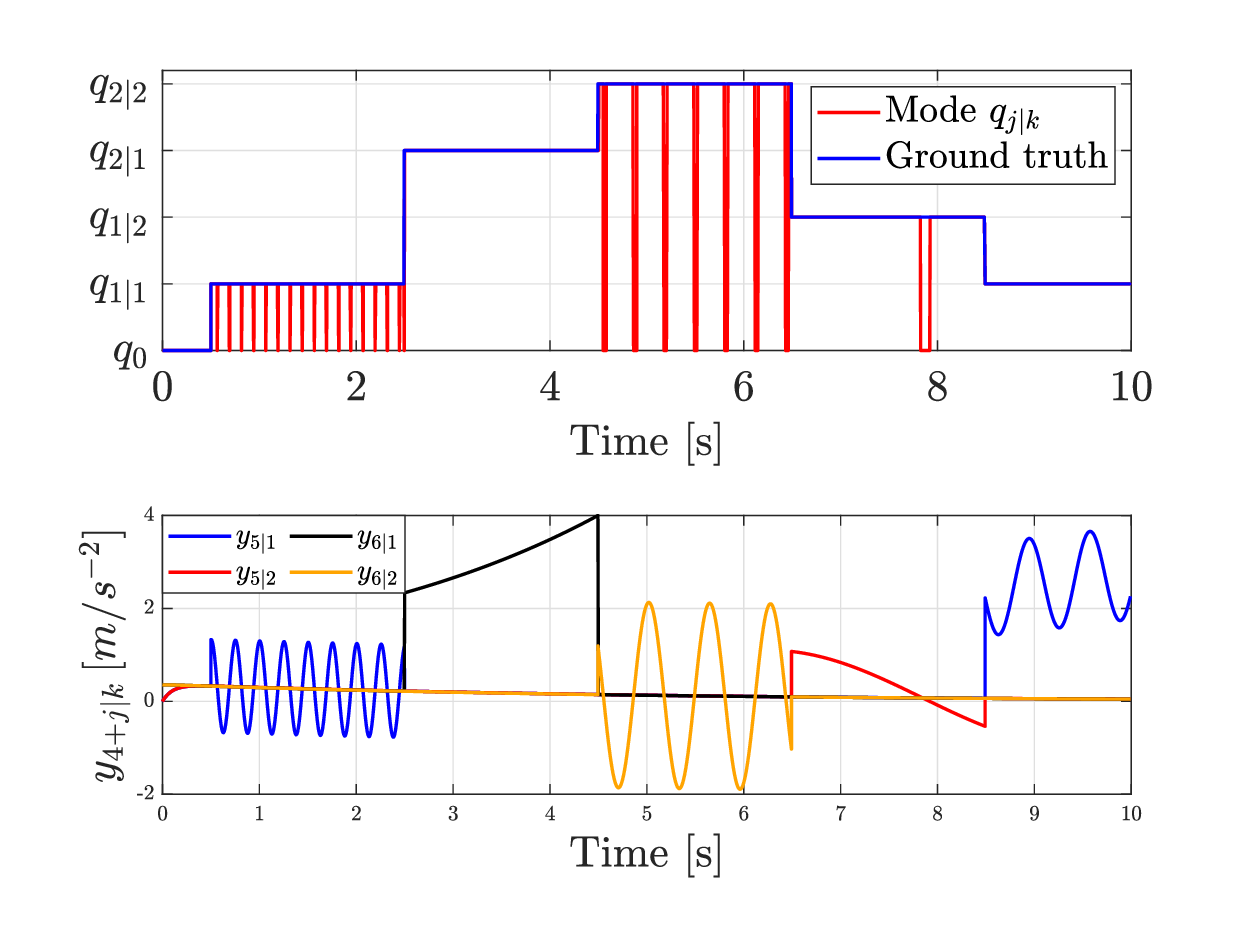}
    \vspace{-7.5mm}
    \caption{Resulting discrete mode $q \in Q$ with the controller output shown in red (indicating the active mode), ground truth mode transitions in blue, and system output $y_{4+j|k}$ for $j,k$ during the FDI attack sequence \eqref{eq:AttSequence}.}
    \centering
    \label{fig:transient_sensAtt}
    \vspace{-0mm}
\end{figure}

Fig.~\ref{fig:transient_sensAtt} shows the discrete mode $q$ alongside the system output $y_{4+j|k}$ for $j,k \in \{1,2\}$. The hybrid controller successfully identifies the compromised sensor as the discrete mode $q$ (red line) follows the ground truth mode transition (blue line). However, occasionally the controller jumps to the healthy mode $q_0$ despite the attack persisting. This occurs when the injected signal causes two corresponding sensor outputs to coincide, triggering the guard $G(q_{1|1}, q_0)$ and the associated reset. In such a case, a reset is necessary because the attack may have stopped. However, the mode immediately jumps back to $q_{1|1}$ once the input $u_{1|1}$ deviates from the others. This behavior is not a limitation but a consequence of the specific attack.

%% file: 6.ConclusionFuturework.tex
\section{CONCLUSIONS AND FUTURE WORKS}\label{sec:Conclusion} \vspace{-3mm}
To enhance the robustness of cooperative driving against cyberattacks, we proposed a real-time hybrid control scheme to detect and mitigate FDI attacks on dynamic CACC using two equivalent controller realizations. This approach preserves closed-loop stability and performance by ensuring that a healthy control input is always applied. Simulation results demonstrate real-time attack detection and mitigation without dwell-time constraints. Although demonstrated on a CACC system, the underlying principle of exploiting equivalent controller realizations for hybrid attack-resilient control can be extended to a broad class of cyber-physical systems.

Future work will examine the robustness of the hybrid switching logic under non-ideal conditions. Sensor noise inherently affects detection and reset accuracy; if the controller states $\rho$ and inputs $u$ do not coincide, the resulting mismatch can cause Zeno behavior, leading to high-frequency switching and potential instability. Additionally, detection and transmission delays may induce late switching, potentially compromising vehicle stability. Finally, it is of interest to determine conditions that minimize the number of required sensors while maintaining detection and control performance.